\documentclass[aps,superscriptaddress,twocolumn,floatfix]{revtex4}%showkeys,
\usepackage{amssymb}
\usepackage{amsmath}
\usepackage{graphicx}
\usepackage{verbatim}
\usepackage[normalem]{ulem}
\usepackage[dvips]{color}
\usepackage{multirow}
\usepackage{dcolumn}                 % Align table columns on decimal point
\usepackage[bookmarksnumbered,bookmarksopen,colorlinks,citecolor=blue,linkcolor=blue]{hyperref} %
\usepackage{CJK}                     % chinese fonts
\usepackage[normalem]{ulem} % \sout{old text} for strikeout
\usepackage[dvips]{color} % For blue in-text comments and additions
\renewcommand\sout{\bgroup \color{red} \ULdepth=-.5ex \ULset}

\usepackage{epstopdf, color}

\def\rpi {$\pi^-/\pi^+$~}
\def\es0{$E_{sym}(\rho_0)$~}
\begin{document}
\begin{CJK*}{GBK}{song}
\title{Beam energy dependence of the relativistic retardation effects of electrical fields on the $\pi^{-}/\pi^{+}$ ratio in heavy-ion collisions}
\author{Gao-Feng Wei}\email[Corresponding author. E-mail: ]{wei.gaofeng@foxmail.com}
\affiliation{School of Mechanical and Material Engineering, Xi'an University of Arts and Sciences, Xi'an 710065, China}
\affiliation{School of Physics and Electronic Science, Guizhou Normal University, Guiyang 550001, China}
\author{Gao-Chan Yong}
\affiliation{Institute of Modern Physics,
Chinese Academy of Sciences, Lanzhou 730000, China}
\author{Li Ou}
\affiliation{College of Physics and Technology, Guangxi Normal University, Guilin 541004, China}
\affiliation{Guangxi Key Laboratory Breeding Base of Nuclear Physics and Technology, Guangxi Normal University, Guilin 541004, China}
\author{Qi-Jun Zhi}
\affiliation{School of Physics and Electronic Science, Guizhou Normal University, Guiyang 550001, China}
\affiliation{Guizhou Provincial Key Laboratory of Radio Astronomy and Data Processing, Guizhou Normal University, Guiyang 550001, China}
\author{Zheng-Wen Long}
\affiliation{College of Physics, Guizhou University, Guiyang 550025, China}

\begin{abstract}
 In this article we investigate the beam energy dependence of relativistic retardation effects of electrical fields on the single and double \rpi ratios in three heavy-ion reactions with an isospin- and momentum-dependent transport model IBUU11. With the beam energy increasing from 200 to 400 MeV/nucleon, effects of the relativistically retarded electrical fields on the \rpi ratio are found to increase gradually from negligibly to considerably significant as expectedly; it is however, the interesting observation is the relativistic retardation effects of electrical fields on the \rpi ratio are becoming gradually insignificant as the beam energy further increasing from 400 to 800 MeV/nucleon. That is to say, with the beam energy increasing, the competition of enhanced  anisotropic features of retarded electrical fields and reduced duration time of the reactions gets effects of the relativistically retarded electrical fields on the \rpi ratio to be maximum around 400 MeV/nucleon. Therefore, the relativistic retardation effects of electrical fields should be carefully considered in heavy-ion collisions at intermediate energy especially around 400 MeV/nucleon when using the \rpi ratio as the probe of nuclear symmetry energy. Moreover, we also investigate the isospin dependence of relativistic retardation effects of electrical fields on the \rpi ratio in two isobar reaction systems of $^{96}$Ru+$^{96}$Ru and $^{96}$Zr+$^{96}$Zr at the beam energies from 200 to 800 MeV/nucleon. It is shown that the relativistic retardation effects of electrical fields on the \rpi ratio are independent of the isospin of reaction. Furthermore, we also examine the double \rpi ratio in reactions of $^{96}$Zr+$^{96}$Zr over $^{96}$Ru+$^{96}$Ru at the beam energies from 200 to 800 MeV/nucleon with the static field and retarded field, respectively. It is shown the double \rpi ratio from two reactions is still an effective observable of symmetry energy without the interference of electrical field due to using the relativistic calculation compared to the nonrelativistic calculation.
\end{abstract}

%\pacs{41.20.-q, %Applied classical electromagnetism
%      25.70.-z, %Low and intermediate energy heavy-ion reactions
%      24.10.Lx, %Monte Carlo simulations (including hadron and parton cascades and string breaking models)
%      21.65.-f  %Nuclear matter
%      25.75.Ld %Collective flow

%      }
%\keywords{Relativistic retardation effect; symmetry energy; heavy-ion reactions, pion production, collective flow}
\maketitle

%\section{Introduction}\label{introduction}
The determination of equation of state (EoS) of isospin asymmetric nucleonic matter has been a longstanding interest in nuclear physics and nuclear astrophysics due to its importance in understanding the properties of radioactive nuclei and evolution of supernova and neutron stars \cite{Steiner05,Lat12,Hor14,Heb15}. As an important tool in terrestrial laboratories, heavy-ion collisions (HICs) with neutron-rich nuclei provide an unique opportunity to form directly the dense and isospin asymmetric nucleonic matter, and thus comparisons of the theoretical simulations of isospin observables with the corresponding experimental measurement enable to extract the EoS of isospin asymmetric nucleonic matter. Actually, many observables have been proposed as the promising isospin signals to detect the isospin ingredients of nuclear effective interactions. These include the isospin fractionation \cite{Muller95,Baran02}, the isospin diffusion \cite{Shi03,LiBA04}, the neutron-proton transverse difference flow \cite{Scalone99}, the neutron-proton correlation function \cite{Chen03a}, the \rpi ratio \cite{BALi02,Gai04}, the $K^{0}/K^{+}$ ratio \cite{Ferini06}, the elliptic flow \cite{BALi01,Dan02}, and the dynamical dipole oscillators \cite{Pie09,Zheng17} etc. However, compared to the achieved significant progress on the EoS of symmetric nucleonic matter \cite{Dan02}, the knowledge on the EoS of isospin asymmetric nucleonic matter is still rather few, especially for the isospin asymmetric nucleonic matter at high baryonic density \cite{ditoro,LCK08,lynch09}.

Presently, the most uncertain part of EoS of isospin asymmetric nucleonic matter is the nuclear symmetry energy at high densities. For example, through comparing the pion observable with the FOPI data \cite{FOPI}, the isospin-dependent Boltzmann-Uehling-Uhlenbeck (IBUU) \cite{Xiao09} model and isospin-dependent Boltzmann-Langevin (IBL) \cite{Xie13} model favor a supersoft prediction for the high-density behaviour of nuclear symmetry energy, but a completely opposite prediction, i.e., a superstiff symmetry energy at high-densities is supported by the Lanzhou quantum molecular dynamics (LQMD) model \cite{Feng10}. While comparing the elliptic flow with the data from the FOPI-LAND collaborators \cite{FOPI-LAND} and/or ASY-EOS experiments \cite{ASY-EOS}, a moderately soft symmetry energy at high densities is concluded from predictions using both the ultrarelativistic quantum molecular dynamics (UrQMD) \cite{Rus11,Trau12,ASY-EOS} model and T\"{u}bingen quantum molecular dynamics (T\"{u}QMD) \cite{Cozma13} model. This is mainly due to the fact that the isovector potential of nuclear effective interactions is much weaker than that of isoscalar one, and thus the corresponding isospin signals are usually interfered by other factors during experimental measurement and theoretical simulations. It is therefore the covariance analysis of these isospin observables \cite{Zhang15} and using some strategies \cite{Tsang17} to maximize the effects of symmetry energy on them are necessary, except for the experimental efforts \cite{FOPI-LAND,FOPI,ASY-EOS,Shane}.
Certainly, to better constrain the high-density behaviour of nuclear symmetry energy, the transport reaction theory communities have been making efforts to compare their codes under controlled conditions to better understand the origins of discrepancies among widely used transport models, and expect to extract reliable information about the EoS of isospin asymmetric matter using these isospin observables from HICs at the intermediate energy \cite{Aichelin89,kolo05,trans1,trans2}.

The ratio \rpi of charged pions is found to be one of the most promising isospin observables, especially concerning the aspects of probing the high-density behavior of nuclear symmetry energy using HICs \cite{BALi02,Gai04,FOPI,Uma98,AMD,Cozma16,Tsang17,Xiao09,Feng10,Xie13}.
Nevertheless, some recent studies on the pion observables have shown clearly that the pion potential \cite{XuKo10,Hong14,Song15,Guo15a,Zhang17,Feng17}, the uncertainty of the $\Delta$ isovector potential \cite{Bao15a,Guo15b}, the neutron-skins of colliding nuclei \cite{Wei14}, and the isospin-dependent high-momentum tails of nucleons induced by the short-range correlations \cite{Ohen14,Ohen15,Bao15b,Yong16}, can all interfere appreciably the sensitivities of \rpi ratio of charged pions in probing the nuclear symmetry energy using HICs. Therefore, before making a final conclusion about the high-density behaviour of nuclear symmetry energy through comparing theoretical simulations of pion observable with the data, one should check all of possible uncertain factors in theoretical simulations of this observable, and thus to get it clean as much as possible in probing the symmetry energy using HICs.

Very recently, we have shown that the relativistically retarded electrical fields created by fast-moving charged particles are anisotropic compared to the isotropic static Coulomb fields as normally used in most of the transport models, and affect significantly the \rpi ratio in a typical reaction of $^{197}$Au+$^{197}$Au collision at a beam energy of 400 MeV/nucleon available in several laboratories \cite{Wei18}. As a subsequent consideration, one naturally assumes that the relativistic retardation effects of electrical fields created by these fast-moving charged particles on the \rpi ratio may increase with increasing the beam energy of HICs. To answer this question precisely, in this article we perform the Au+Au collisions at the beam energies from 200 to 800 MeV/nucleon to show the beam energy dependence of relativistic retardation effects of electrical field on the \rpi ratio. As one expected, with the beam energy increasing from 200 to 400 MeV/nucleon, effects of the relativistically retarded electrical fields on the \rpi ratio are increasing gradually from negligibly to considerably significant; it is however, the interesting observation is the relativistic retardation effects of electrical fields on the \rpi ratio are becoming gradually insignificant as the beam energy further increasing from 400 to 800 MeV/nucleon due to the competition of reduced duration time of the reaction and the enhanced anisotropic features of retarded electrical fields. Moreover, we also investigate the isospin dependence of relativistic retardation effects of electrical fields on the \rpi ratio in two isobar reaction systems of $^{96}$Ru+$^{96}$Ru and $^{96}$Zr+$^{96}$Zr at the beam energies from 200 to 800 MeV/nucleon. It is shown the relativistic retardation effects of electrical fields on the \rpi ratio are independent of the isospin of reaction. Furthermore, we also examine the double \rpi ratio from reactions of $^{96}$Zr+$^{96}$Zr over $^{96}$Ru+$^{96}$Ru at the beam energies from 200 to 800 MeV/ncucleon with the static field and retarded field, respectively. It is shown the double \rpi ratio in two reactions can not only eliminate effectively effects of the electrical fields due to with the relativistic calculation compared to the nonrelativistic calculation but also sustain the sensitivities to symmetry energy, and thus is still an effective observable of symmetry energy in HICs at intermediate energy.
%\section{The model}\label{model}

Similar to our recent work \cite{Wei18}, the present study is carried out within the isospin- and momentum-dependent Boltzmann-Uehling-Uhlenbeck transport model \cite{Das03,IBUU} of IBUU11 version. In this model, the nuclear mean-field interaction is expressed as
\begin{eqnarray}
U(\rho,\delta ,\vec{p},\tau ) &=&A_{u}(x)\frac{\rho _{-\tau }}{\rho _{0}}%
+A_{l}(x)\frac{\rho _{\tau }}{\rho _{0}}+\frac{B}{2}{\big(}\frac{2\rho_{\tau} }{\rho _{0}}{\big)}^{\sigma }(1-x)  \notag \\
&+&\frac{2B}{%
\sigma +1}{\big(}\frac{\rho}{\rho _{0}}{\big)}^{\sigma }(1+x)\frac{\rho_{-\tau}}{\rho}{\big[}1+(\sigma-1)\frac{\rho_{\tau}}{\rho}{\big]}
\notag \\
&+&\frac{2C_{\tau ,\tau }}{\rho _{0}}\int d^{3}p^{\prime }\frac{f_{\tau }(%
\vec{p}^{\prime })}{1+(\vec{p}-\vec{p}^{\prime })^{2}/\Lambda ^{2}}
\notag \\
&+&\frac{2C_{\tau ,-\tau }}{\rho _{0}}\int d^{3}p^{\prime }\frac{f_{-\tau }(%
\vec{p}^{\prime })}{1+(\vec{p}-\vec{p}^{\prime })^{2}/\Lambda ^{2}},
\label{MDIU}
\end{eqnarray}%
and the parameters $A_{l}(x)$, $A_{u}(x)$ are expressed in the forms of
\begin{eqnarray}
A_{l}(x)&=&A_{l0} - \frac{2B}{\sigma+1}\big{[}\frac{(1-x)}{4}\sigma(\sigma+1)-\frac{1+x}{2}\big{]},  \\
A_{u}(x)&=&A_{u0} + \frac{2B}{\sigma+1}\big{[}\frac{(1-x)}{4}\sigma(\sigma+1)-\frac{1+x}{2}\big{]}.
\end{eqnarray}
It should be mentioned that in the IBUU11 version of this model we have updated the expression of nuclear mean-field interaction and/or readjusted the corresponding parameters to consider more accurately the spin-isospin dependence of in-medium effective many-body forces by distinguishing the density dependences of $nn$, $pp$ and $np$ interactions in the effective three-body force term \cite{CXu10,Chen14b}, as well as the high-momentum behaviors of the nucleon optical potential extracted from nucleon-nucleus scattering experiments \cite{LXH13}. The values of these parameters used in this study are determined as $A_{l0}(x)$ = -76.963 MeV, $A_{u0}(x)$ = - 56.963 MeV, $B$= 141.963 MeV, $C_{\tau ,\tau }$= -57.209 MeV, $C_{\tau ,-\tau }$= -102.979 MeV, $\sigma$= 1.2652 and $\Lambda $= 2.424$p_{f0}$ where $p_{f0}$ is the nucleon Fermi momentum in symmetric nuclear matter at normal density $\rho_{0}$=0.16 fm$^{-3}$. Using above single particle potential, one can derive the symmetry energy for a specific parameter $x$, which is introduced to mimic the different forms of symmetry energy predicted by various many-body theories without changing any property of symmetric nuclear matter and the value of symmetry energy $E_{sym}(\rho_{0})$=30.0 MeV at saturation density.

%\section{Results and Discussions}\label{results}
On the other hand, the relativistically retarded electric fields created by fast-moving charged particles in HICs are calculated according to the well-known Li\'{e}nard-Wiechert expression
\begin{equation}\label{electrical field}
e\vec{E}(\vec{r},t)=\frac{e^2}{4\pi \varepsilon_{0}}
\sum_{n}Z_{n}\frac{c^2-v^{2}_{n}}{(cR_{n}-\vec{R}_{n}\cdot \vec{v}_{n})^3}(c\vec{R}_{n}-R_{n}\vec{v}_{n}),
\end{equation}
where $Z_{n}$ is the charge number of the $n$th particle, $\vec{R}_{n}=\vec{r}-\vec{r}_{n}$ is the position of the field point $\vec{r}$ relative to the source
point $\vec{r}_{n}$ where the $n$th particle is moving with velocity $\vec{v}_{n}$ at the retarded time of $t_{n}=t-|\vec{r}-\vec{r}_{n}|/c$.
It is necessary to mention that to calculate the retarded electric fields $e\vec{E}(\vec{r},t)$, the phase space histories of all charged particles before the moment $t$ have to be saved in transport model simulations. Moreover,
a pre-collision phase space history for all nucleons is made assuming that they are frozen in the projectile and target moving along their Coulomb trajectories. More technical details about calculating the $e\vec{E}(\vec{r},t)$ and numerical checks can be found in Ref. \cite{Ou11}.
Whereas, in the nonrelativistic limit $v_{n}$$\ll$$c$,
Eq.(\ref{electrical field}) naturally reduces to the static Coulomb field of the form
\begin{equation}\label{Coulomb field}
e\vec{E}(\vec{r},t)=\frac{e^2}{4\pi \varepsilon_{0}}
\sum_{n}Z_{n}\frac{\vec{R}_{n}}{R_{n}^{3}}.
\end{equation}
\begin{figure}[h]
\centerline{\includegraphics[width=1.1\columnwidth]{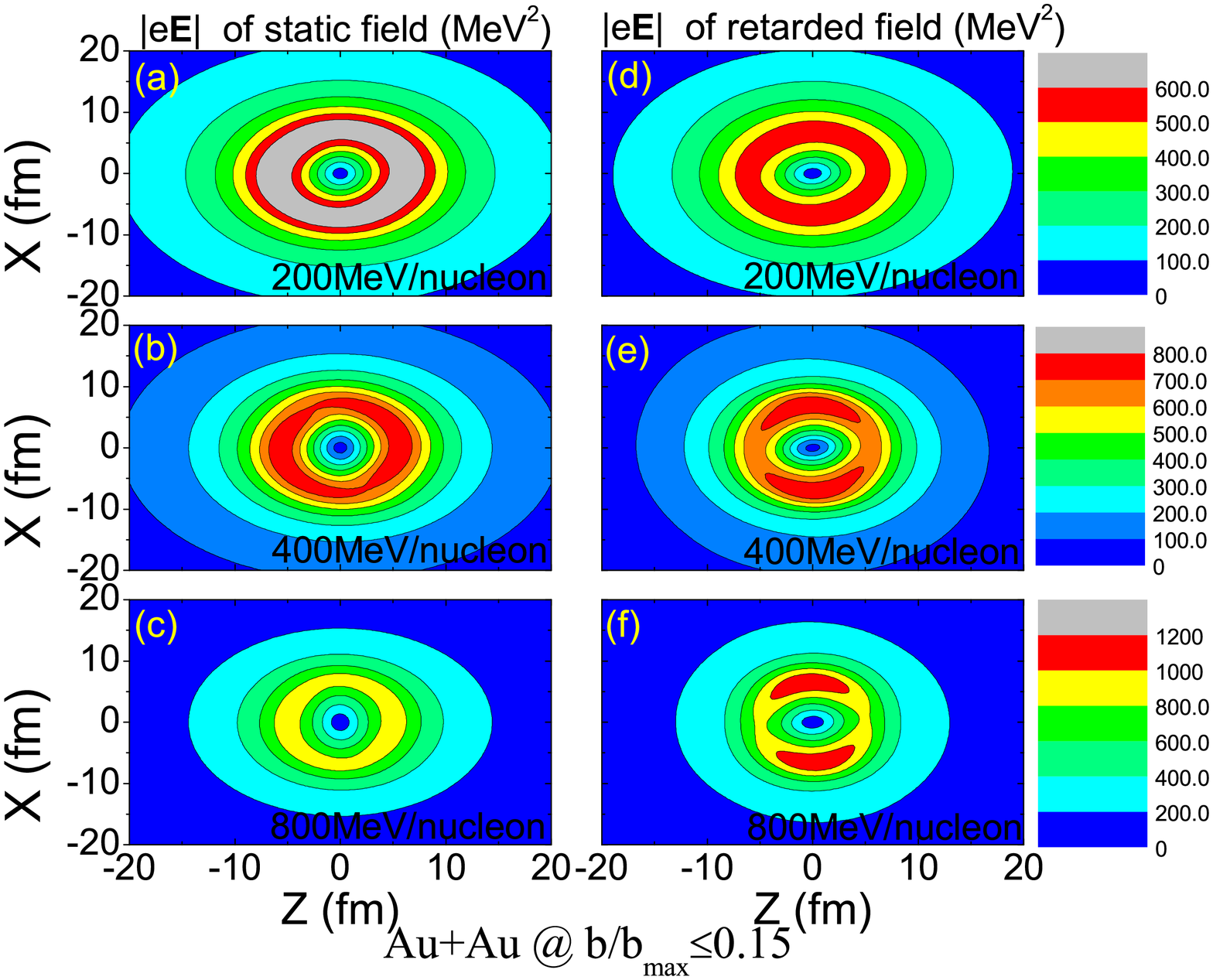}}
\vspace{0.3cm}
\centerline{\includegraphics[width=1.1\columnwidth]{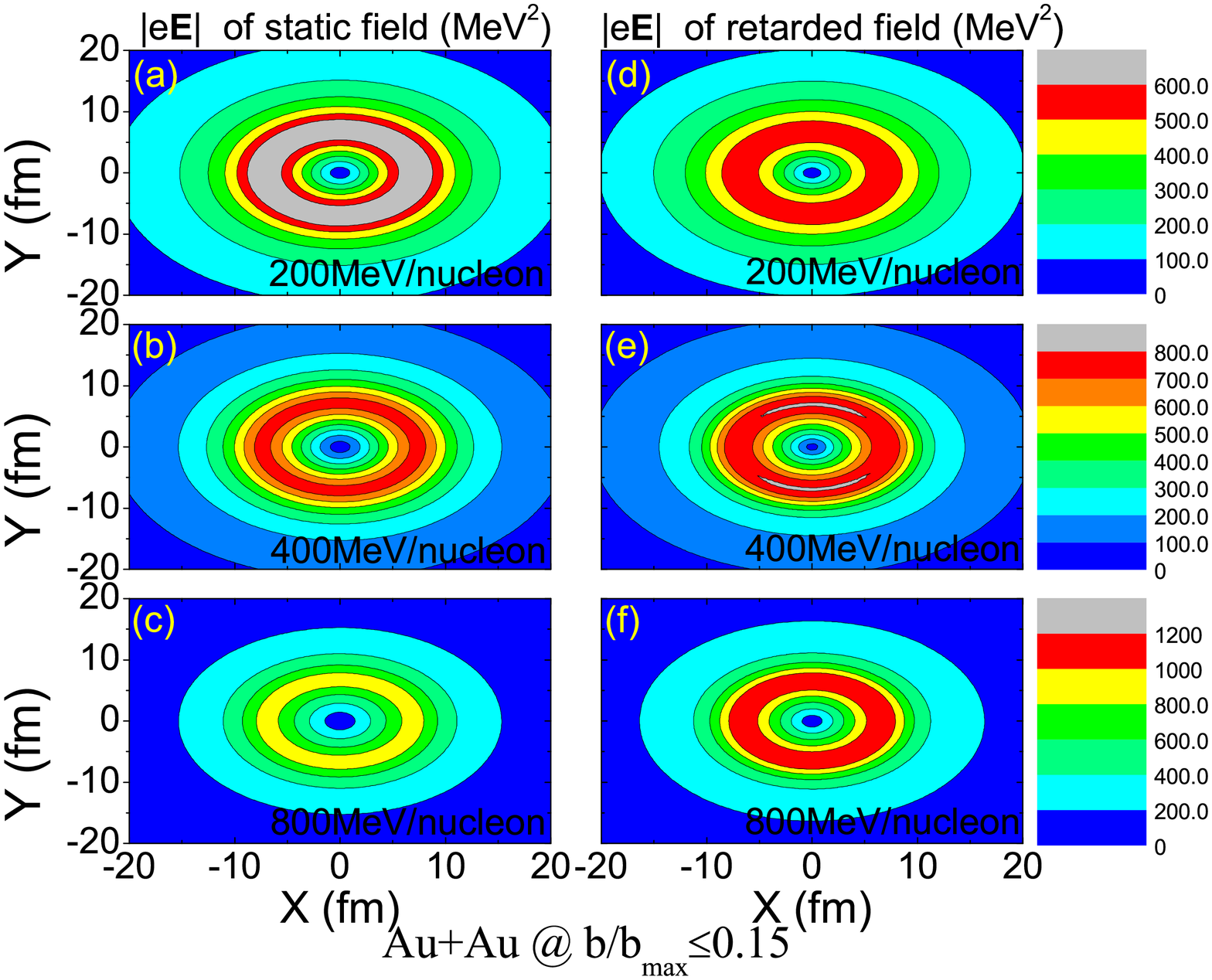}}
\caption{(Color online) Contours of the electric fields in the X-o-Z reaction plane (upper) and X-o-Y plane (lower) at the maximum compression stage in central Au+Au collisions at the beam energies of 200, 400 and 800 MeV/nucleon. The panels (a), (b) and (c) are for the static fields while the panels (d), (e) and (f) are for the retarded fields.} \label{field}
\end{figure}

As has been shown in our recent work \cite{Wei18} in Au+Au collision at 400 MeV/nucleon the maximum difference between the retarded and static fields is located in the reaction (X-o-Z) plane at the maximum compression stage. Therefore, we plot in the upper window of Fig. \ref{field} the strength $|e\bold{E}|$ contours of the static and retarded electric fields at the maximum compression stage in the X-o-Z reaction plane at three beam energies of 200, 400 and 800 MeV/nucleon. It is obvious to see the most important differences between them are the non-isotropic nature of the retarded electrical fields created by fast-moving charged particles, i.e., transversely enhanced by a factor $\gamma=1/(1-\beta^2)^{1/2}$ while longitudinally reduced by $1/\gamma^2$. Certainly, the retardation effect depends on the reduced velocity $\beta=v/c$ of charged particles and thus depends on the beam energies. In fact, in the relativistic case, all charged particles with velocity $\vec{v}_{n}$ at the retarded time $t_n$ contribute to the $e\vec{E}(\vec{r},t)$ at the instant $t$ and location $\vec{r}$; whereas in the nonrelativistic case, those charged particles contribute to the $e\vec{E}(\vec{r},t)$ just at the same moment $t$. More specifically, due to the longer retardation time of the reaction at the lower beam energy compared to the case at the higher beam energy, the retarded electrical field shows more weaker strength than those of static field at 200 MeV/nucleon; whereas in the higher beam energy, the more stronger Lorentz contraction gets the retarded electrical fields showing more clearly non-isotropic feature at 800 MeV/nucleon. Whereas in the X-o-Y plane which is perpendicular to the beam direction and thus has the symmetric electrical components in X and Y directions, it is therefore the feature of non-isotropic is invisible in retarded electrical fields as shown in the lower window of Fig. \ref{field}. Certainly, due to the electrical fields in the relativistic calculations stem from all the charged particles at the retarded time $t_{n}$, whereas those in the nonrelativistic calculations just from those of charged particles at the same moment $t$, it is thus the weaker (stronger) strength in retarded field than static field at lower (higher) beam energy of 200 (800) MeV/nucleon is observed in X-o-Y plane as shown in the lower window of Fig. \ref{field}.
\begin{figure}[th]
\centerline{\includegraphics[width=1.1\columnwidth]{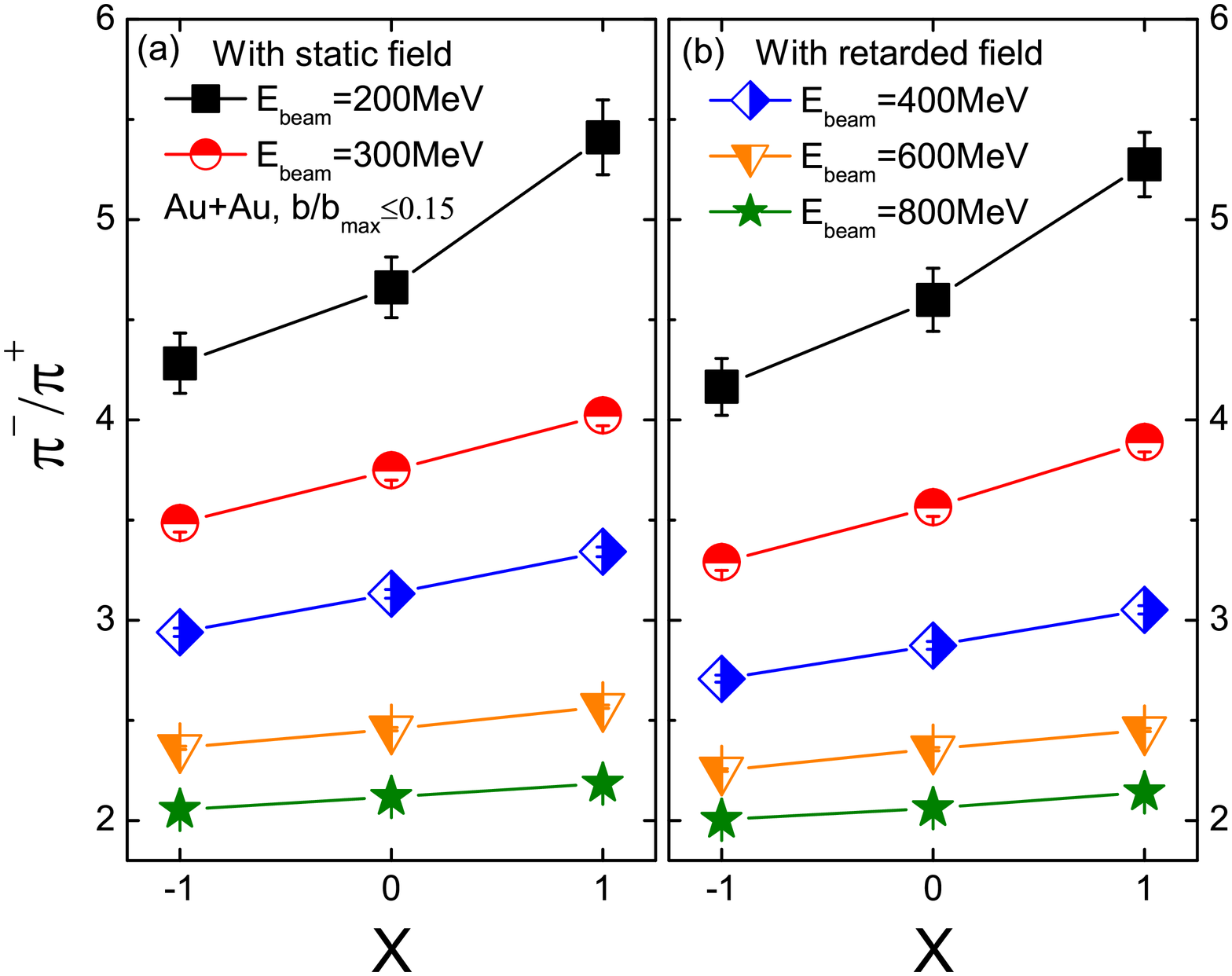}}
\centerline{\includegraphics[width=1.1\columnwidth]{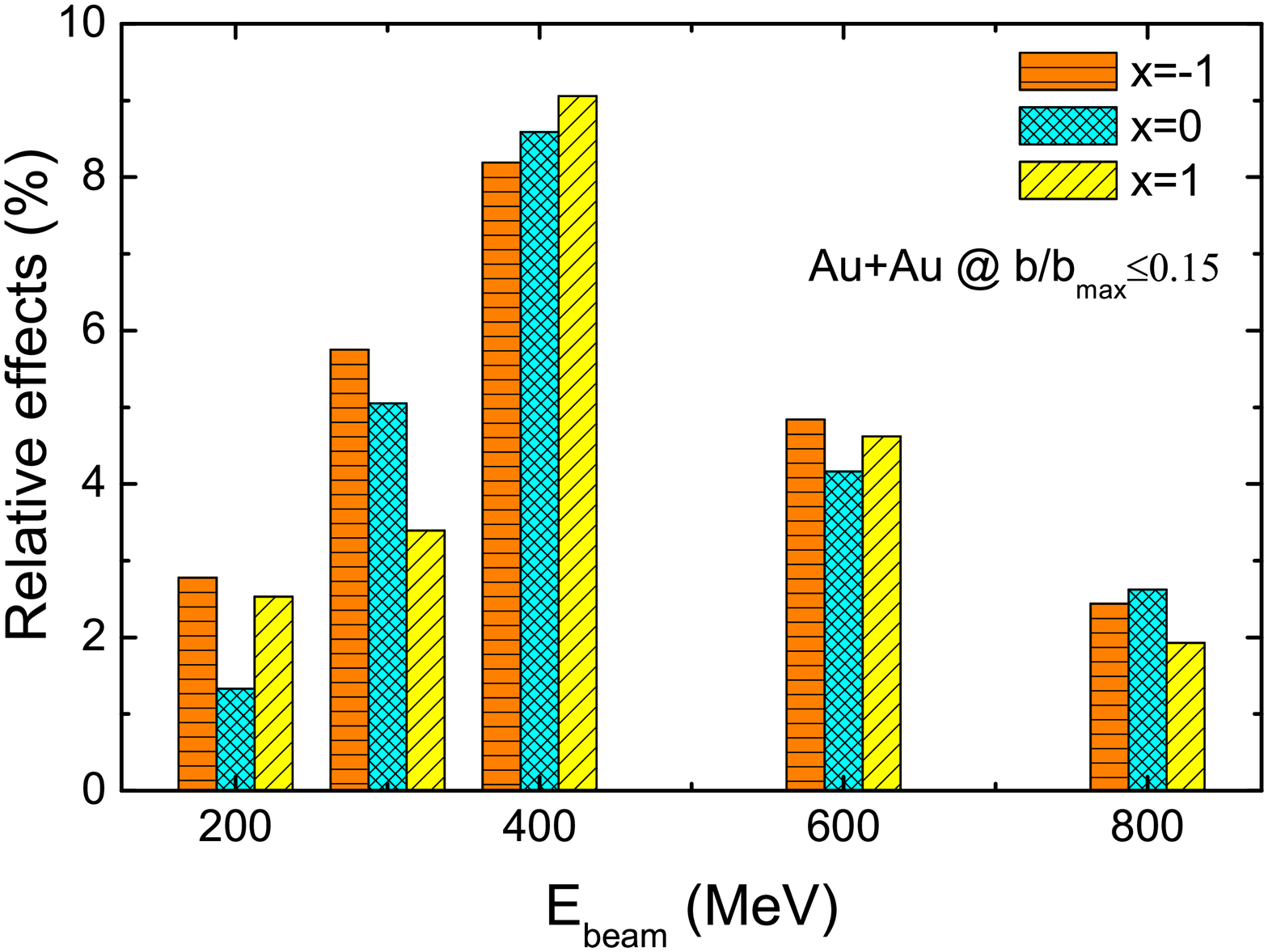}}
\caption{(Color online) Upper: The final \rpi ratio in central Au+Au collisions as a function of beam energy with three symmetry energy settings from the hard of $x$=-1 to soft of $x$=1. Lower: The relative effects of relativistically retarded electrical fields on the final $\pi^{-}/\pi^{+}$ ratio shown in the same reactions. } \label{rpi}
\end{figure}

Now, we turn to the beam energy dependence of relativistic retardation effects of electrical fields on the \rpi ratio. Shown in upper window of Fig. \ref{rpi} is the final \rpi ratio in central Au+Au collisions at the beam energies from 200 to 800 MeV/nucleon with the symmetry energy settings from the hard of $x$=-1 to soft of $x$=1. Firstly, consistent with the established systematics for pion production \cite{FOPI},  the \rpi ratio is decreasing with the beam energy increasing and shows more sensitivities to the density dependent nuclear symmetry energy at the lower beam energy especially around and below the pion production threshold regardless how the electrical fields are calculated. Secondly, it can be found that the final \rpi ratio is smaller on the whole with the retarded electric fields than those with the static ones in collisions at all beam energies studied here. Thirdly, with the beam energy increasing from 200 to 400 MeV/nucleon, the relativistic retardation effects of electrical fields on \rpi ratio are found to increase gradually from negligibly to considerably significant as expectedly; it is however, one can also see the relativistic retardation effects of electrical fields on \rpi ratio are becoming gradually insignificant as the beam energy further increasing from 400 to 800 MeV/nucleon. More specifically, the relative effect of retarded electrical fields on the ratio \rpi of charged pions can reach maximally about 8-9\% at 400 MeV/nucleon but less than about 3\% at both 200 MeV/nuclon and 800 MeV/nucleon as shown in lower window of Fig. \ref{rpi}.  Nevertheless, the more prominent relativistic effects of charged particles moving with higher speed get one naturally assuming that the effects of relativistically retarded electrical fields on \rpi ratio are increasing as the beam energy increasing. Therefore, how to understand this interesting observation is the main task we shall discuss in the following.

\begin{figure}[h]
\centerline{\includegraphics[width=1.1\columnwidth]{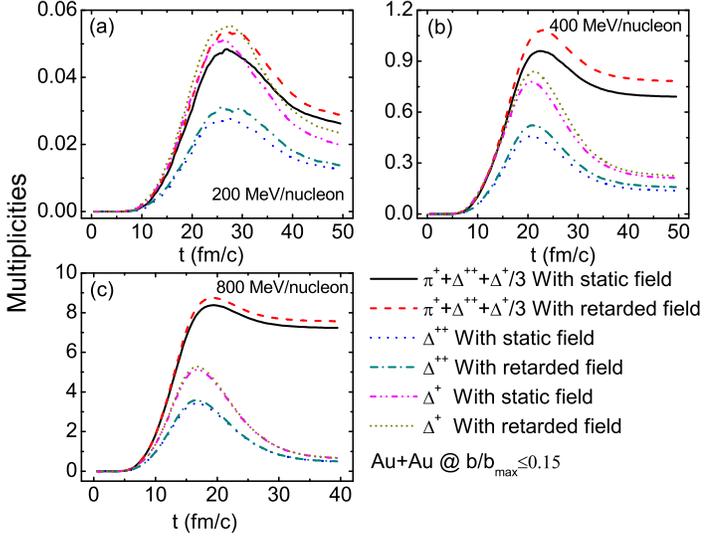}}
%\vspace{0.3cm}
%\centerline{\includegraphics[width=1.0\columnwidth]{fig7.eps}}
%\vspace{0.3cm}
%\centerline{\includegraphics[width=0.86\columnwidth]{fig8.eps}}
\caption{(Color online) Evolution of dynamical multiplicities of $\pi^{+}$ (i.e., $\pi^{+}+\Delta^{++}+\frac{1}{3}\Delta^{+}$), $\Delta^{++}$ and $\Delta^{+}$ resonances in central Au+Au collisions at the beam energies of 200, 400 and 800 MeV/nucleon, respectively. } \label{mul-plus}
\end{figure}

\begin{figure}[h]
\centerline{\includegraphics[width=1.1\columnwidth]{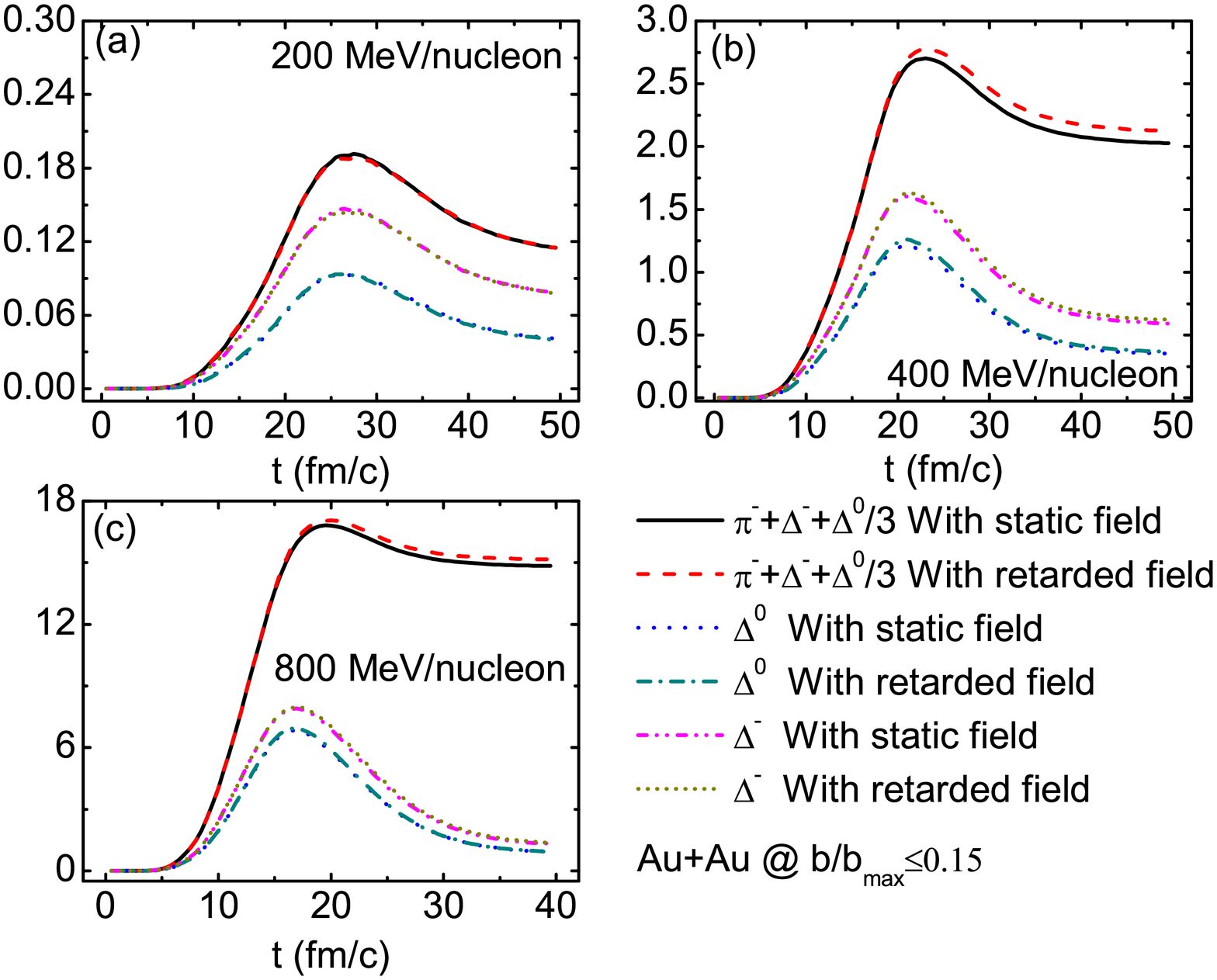}}
%\vspace{0.3cm}
%\centerline{\includegraphics[width=1.0\columnwidth]{fig7.eps}}
%\vspace{0.3cm}
%\centerline{\includegraphics[width=0.86\columnwidth]{fig8.eps}}
\caption{(Color online) Evolution of dynamical multiplicities of $\pi^{-}$ (i.e, $\pi^{-}+\Delta^{-}+\frac{1}{3}\Delta^{0}$), $\Delta^{-}$ and $\Delta^{0}$ resonances in central Au+Au collisions at the beam energies of 200, 400 and 800 MeV/nucleon, respectively. } \label{mul-minus}
\end{figure}

For pion production in HICs at intermediate energies, almost all of pions are from the decay of $\Delta$(1232) resonances. As a result, the multiplicities of dynamical $\pi^{-}$ and $\pi^{+}$ during reactions are determined as $\pi^{-}+\Delta^{-}+\frac{1}{3}\Delta^{0}$ and $\pi^{+}+\Delta^{++}+\frac{1}{3}\Delta^{+}$ according to the decay mechanism of $\Delta$(1232) resonances. Of course, all the $\Delta$ resonances will eventually decay into nucleons and pions at the end of reaction. Naturally, the $\pi^{+}+\Delta^{++}+\frac{1}{3}\Delta^{+}$ and $\pi^{-}+\Delta^{-}+\frac{1}{3}\Delta^{0}$ will become $\pi^{+}$ and $\pi^{-}$ mesons at the end of reactions. However, to better understand how the retarded electrical fields affect the $\Delta$(1232) resonances during reactions and thus the multiplicities of charged pions and their ratio \rpi at the end of reactions, it is necessary to trace the dynamical production of $\Delta$(1232) resonances and their contributions to the multiplicities of dynamical $\pi^{-}$ and $\pi^{+}$. Shown in Fig. \ref{mul-plus} and Fig. \ref{mul-minus} are the  dynamical multiplicities of $\pi^{+}$ (i.e., $\pi^{+}+\Delta^{++}+\frac{1}{3}\Delta^{+}$) and $\pi^{-}$ (i.e., $\pi^{-}+\Delta^{-}+\frac{1}{3}\Delta^{0}$) and the corresponding $\Delta$(1232) resonances with different charge states in central Au+Au collisions at three beam energies of 200, 400 and 800 MeV/nucleon. Obviously, regardless of the $\Delta^{++}$ or $\Delta^{+}$ resonances, the retarded fields can increase their multiplicities compared to those with the static field. Naturally, the corresponding dynamical multiplicities of $\pi^{+}$ (i.e., $\pi^{+}+\Delta^{++}+\frac{1}{3}\Delta^{+}$) are larger with the retarded field than those with the static field. However, the increasing of multiplicities of $\Delta^{-}$ and $\Delta^{0}$ resonances is very little and only obviously visible at the beam energy of 400 MeV/nucleon. As a result, the corresponding dynamical multiplicities of $\pi^{-}$ (i.e., $\pi^{-}+\Delta^{-}+\frac{1}{3}\Delta^{0}$) are tinily larger with the retarded field than those with the static field. This is the reason why we see in Fig. \ref{rpi} the ratio \rpi of charged pions is smaller on the whole with the retarded electrical fields than those with the static ones. On the other hand, while the non-isotropic feature of retarded electrical fields and thus their instantaneous effects on both $\Delta^{++}$ and $\Delta^{+}$ resonances at the compression stage are increasing with the beam energy increasing, the corresponding duration time of compression stage of the reaction gets to be decreased. Consequently, the competition of these two factors gets the effects of retarded fields on the final multiplicities of $\pi^{+}$ and thus the \rpi ratio to be maximum around 400 MeV/nucleon.

\begin{figure}[th]
\centerline{\includegraphics[width=1.1\columnwidth]{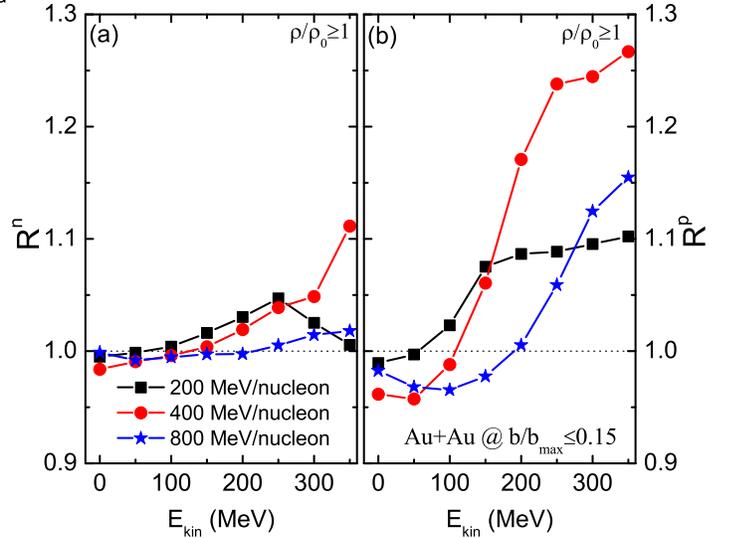}}
%\centerline{\includegraphics[width=1.04\columnwidth]{fig1b.eps}}
\caption{(Color online) The ratio $R^{n}$ for neutrons and $R^{p}$ for protons at supra-saturation densities at the
maximum compression stage as a function of nucleon kinetic energy in central Au+Au collisions at the beam energies of 200, 400 and 800 MeV/nucleon, respectively.} \label{rnp}
\end{figure}

Certainly, one may ask why and how the multiplicities of the $\Delta$(1232) resonances especially the $\Delta^{++}$ and $\Delta^{+}$ resonances are affected by the retarded fields. To answer this question, we investigate the relative change in nucleon kinetic energy distributions due to using the retarded electrical fields compared to the static ones. For this purpose, we examine the ratio
\begin{equation}
R^{i}=\frac{{\rm Number}(i)_{\rm{R}}}{{\rm Number}(i)_{\rm{S}}},~~~i\equiv{\rm{neutron~or~proton}}
\end{equation}\label{R}\\
of nucleons with local densities higher than $\rho_0$ at the maximum compression stage in the Au+Au reactions with the retarded (R) and static (S) electrical fields. Shown in Fig. \ref{rnp} are the $R^{n}$ and $R^{p}$ as a function of nucleon kinetic energy in Au+Au collisions at three beam energies of 200, 400 and 800 MeV/nucleon. Obviously, for the energetic nucleons above a certain kinetic energy depending on the beam energy of reactions, the values of both $R^{n}$ and $R^{p}$ are larger than 1, indicating that the retarded fields  increase (decrease) the number of high (low) energy nucleons. These increased energetic protons (neutrons) naturally increase some $pp$ ($nn$) inelastic collisions and thus increase the multiplicities of $\Delta^{++}$ and $\Delta^{+}$ ($\Delta^{-}$ and $\Delta^{0}$) through $p+p\rightarrow\Delta^{++}+n$ and $p+p\rightarrow\Delta^{+}+p$ ($n+n\rightarrow\Delta^{-}+p$ and $n+n\rightarrow\Delta^{0}+n$) reaction channels, and thus leads to more $\pi^{+}$ ($\pi^{-}$) mesons through the decay of these resonances. Certainly, one may suspect why the kinetic energy distribution of neutrons is also affected slightly by the retarded electrical fields as shown in Fig. \ref{rnp}. Actually, the neutrons are not affected directly by the electrical fields, it is however, secondary collisions between neutrons and energetic protons provide the chance for these neutrons increasing their kinetic energies at a lower level compared to those of protons. Nevertheless, as the secondary effects the increased $\Delta^{-}$ and $\Delta^{0}$ and thus the $\pi^{-}$ at the final reaction stage is relative fewer than those of $\pi^{+}$, and thus the final \rpi ratio is smaller on the whole with the retarded fields than those with the static ones. Again, with the beam energy increasing, the competition  between the enhanced non-isotropic feature of retarded electrical fields and reduced duration time of compression stage of the reaction gets the relativistic retardation effects of electrical field on nucleon kinetic energy distribution to be maximum around 400 MeV/nucleon as shown in Fig. \ref{rnp}.
\begin{figure}[th]
\centerline{\includegraphics[width=1.1\columnwidth]{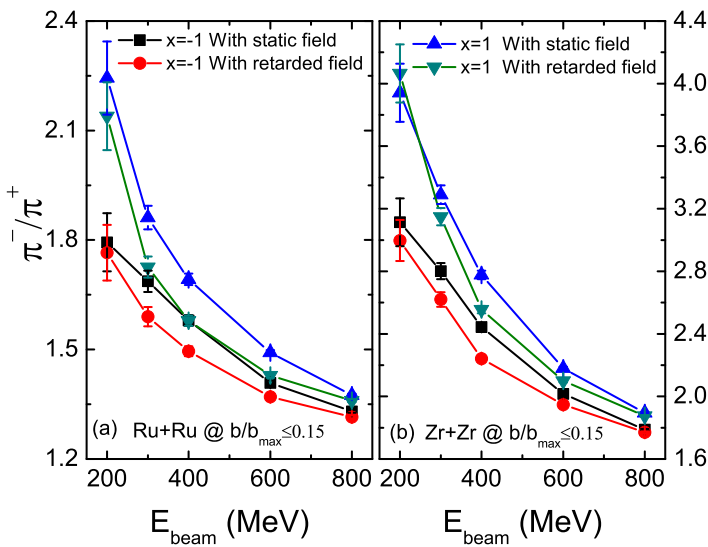}}
%\centerline{\includegraphics[width=1.1\columnwidth]{fig5.eps}}
%\centerline{\includegraphics[width=1.0\columnwidth]{fig6b.eps}}
\caption{(Color online) The final \rpi ratio in two isobar reaction systems of $^{96}$Ru+$^{96}$Ru (a) and $^{96}$Zr+$^{96}$Zr (b) at the beam energies from 200 to 800 MeV/nucleon. The hard symmetry energy with parameter $x$=-1 and the soft one with parameter $x$=1 are used. } \label{iso}
\end{figure}
\begin{figure}[th]
\centerline{\includegraphics[width=1.1\columnwidth]{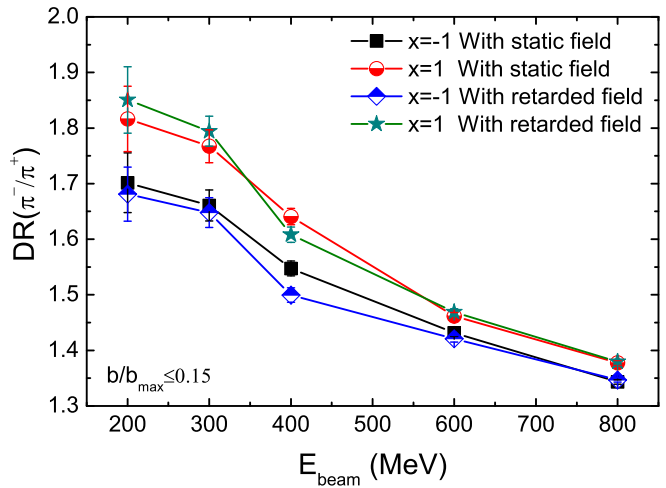}}
%\centerline{\includegraphics[width=1.1\columnwidth]{fig5.eps}}
%\centerline{\includegraphics[width=1.0\columnwidth]{fig6b.eps}}
\caption{(Color online) The double $\pi^{-}/\pi^{+}$ ratio (i.e., DR($\pi^{-}/\pi^{+}$)) of reactions $^{96}$Zr+$^{96}$Zr over $^{96}$Ru+$^{96}$Ru at the beam energies from 200 to 800 MeV/nucleon with the static field and retarded field, respectively. } \label{DR}
\end{figure}

Next, we investigate the isospin dependence of relativistic retardation effects of electrical fields on the \rpi ratio. For this purpose, we show in Fig. \ref{iso} the final \rpi ratio in two isobar reaction systems of $^{96}$Ru+$^{96}$Ru and $^{96}$Zr+$^{96}$Zr at the beam energies from 200 to 800 MeV/nucleon. Again, similar to above in Au+Au reaction, two observations can be found in reactions of $^{96}$Ru+$^{96}$Ru and $^{96}$Zr+$^{96}$Zr. The first is the \rpi ratio shows more sensitivities to the density dependent symmetry energy at the lower beam energy especially around and below the pion production threshold, regardless how the electrical fields are calculated. The second is the influences of relativistically retarded electrical fields on \rpi ratio are more apparent around 400 MeV/nucleon. Moreover, it is seen that the retardation effects of electrical fields on \rpi ratio are independent of the isospin of reaction.

Actually, the relativistic retardation effects of electrical field on the ratio \rpi of charged pions are not desired seen when using this ratio as the probe of symmetry energy in HICs since a key step in determining the nuclear symmetry energy is the determination of experimental observables which can be serve as clean and sensitive probes \cite{Dan98,Wei15}. Moreover, considering the fact that the isovector potentials are generally smaller than the isoscalar potential at the same nuclear density, therefore, the search of experimental observables or their combination which have the maximum effects of isovector potentials without other interferences, such as the Coulomb effects and/or isoscalar effects etc., is a crucial task in probing the nuclear symmetry energy using HICs. The double ratio of two reactions as the candidate of such this kind observables, such as double $n/p$ ratio \cite{Fam06,LiBA06} and double \rpi ratio \cite{Yong06}, is naturally expected to disentangle the effects of symmetry energy from those of electrical fields due to using the relativistic calculation compared to the nonrelativistic calculation, because the double ratio of two reactions has the advantages of reducing the influence of both the Coulomb fields and the systematic errors etc. \cite{Fam06}. Shown in Fig. \ref{DR} is the double \rpi ratio of reactions $^{96}$Zr+$^{96}$Zr over $^{96}$Ru+$^{96}$Ru at the beam energies from 200 to 800 MeV/nucleon with the static field and retarded field, respectively. It is seen the double \rpi ratio can indeed eliminate effectively the effects of electrical fields due to using the relativistic calculation compared to the nonrelativistic calculation, but also sustain the sensitivities to symmetry energy. It is thus can be concluded the double \rpi ratio in two reactions is still an effective observable of symmetry energy in HICs at intermediate energy.
%\section{Summary}\label{summary}

In summary, we have investigated the beam energy depndence of relativistic retardation effects of electrical fields on the single and double \rpi ratios in three heavy-ion reactions. It is shown the relativistic retardation effects of electrical fields on \rpi ratio are increasing firstly and then decreasing gradually with the beam energy increasing from 200 to 800 MeV/nucleon. Specifically, the relativistic retardation effects of electrical fields on the \rpi ratio are found to be maximum at the beam energy around 400 MeV/nucleon due to the competition of anisotropic features of retarded electrical fields and duration time of the reactions. Therefore, the relativistic retardation effects of electrical fields should be carefully considered in HICs at intermediate energy especially around 400 MeV/nucleon when using the \rpi ratio as the probe of nuclear symmetry energy. Moreover, we have also studied the isospin dependence of relativistic retardation effects of electrical fields on the ratio \rpi of charged pions in two isobar reactions of $^{96}$Ru+$^{96}$Ru and $^{96}$Zr+$^{96}$Zr at the beam energies from 200 to 800 MeV/nucleon. It is found that the relativistic retardation effects of electrical fields on the \rpi ratio are independent of the isopin of reaction. Furthermore, we found that the double \rpi ratio of reactions $^{96}$Zr+$^{96}$Zr over $^{96}$Ru+$^{96}$Ru at the beam energies from 200 to 800 MeV/nucleon can not only eliminate effectively the effects of electrical fields due to using the relativistic calculation compared to the nonrelativistic calculation but also can sustain the sensitivities to symmetry energy, and thus can still be as an effective observable of nuclear symmetry energy in HICs at intermediate energy.
Before ending this work, we give an useful remark here. In full covariant transport theories, it has been confirmed that the relativistic mechanisms related to the covariant nature of the nuclear fields can also contribute to the isovector channel and thus affect the sensitivities of collective flows of nucleons in probing the symmetry energy using HICs \cite{Greco03}. Therefore, it is also deserved to study how these relativistic mechanisms related to the covariant nature of the nuclear fields affect the sensitivities of pion observable in probing the nuclear symmetry energy in HICs at intermediate energies.

\section*{\textbf{Acknowledgements}}
G. F. Wei would like to thank Prof. Bao-An Li for helpful discussions. This work is supported in part by the National Natural Science Foundation of China under grant
Nos.11405128, 11465006, 11365004, 11565010, 11775275, U1731218, and the foundation of Guangxi innovative team and distinguished scholar in institutions of higher education and the Natural Science Foundation of Guangxi province under grant No.2016GXNSFFA380001, and the innovation talent team (Grant No.(2015)4015)  and the high level  talents (Grant No.(2016)-4008)) of  the Guizhou Provincial Science and Technology Department.

\end{CJK*}

\end{document}